\documentclass[preprint,nofootinbib,amsmath,amssymb,aps,preprintnumbers,superscriptaddress]{revtex4-1}

\usepackage{bm}
\usepackage{braket}
\usepackage{cases} 
\usepackage{color}
\usepackage{upgreek}
\usepackage{ulem}
\usepackage{comment}
\usepackage{empheq}
\usepackage{float,orcidlink}

\usepackage{hyperref}
\hypersetup{colorlinks=true, allcolors=blue}

\newcommand{\ito}[1]{\textcolor{red}{#1}}

\newcommand{\mnras}{Mon. Not. Roy. Astron. Soc. }
\newcommand{\aap}{Astron. Astrophys. }

\begin{document}

\preprint{KOBE-COSMO-26-03}

\title{Gravitational waves from gaps of neutron stars}

\author{Akira Dohi \orcidlink{0000-0001-8726-5762}}
\email[]{dohi@riken.jp}
\affiliation{Astrophysical Big-Bang Laboratory (ABBL), Pioneering Research Institute (PRI), RIKEN, Wako, Saitama 351-0198, Japan}
\affiliation{Interdisciplinary Theoretical and Mathematical Sciences Program (iTHEMS), RIKEN, Wako, Saitama 351-0198, Japan}

\author{Asuka Ito \orcidlink{0000-0001-8156-0082}}
\email[]{asuka@phys.sci.kobe-u.ac.jp}
\affiliation{Department of Physics, Kobe University, Kobe 657-8501, Japan}

\author{Shota Kisaka \orcidlink{0000-0002-2498-1937}}
\email[]{kisaka@hiroshima-u.ac.jp}
\affiliation{Physics Program, Graduate School of Advanced Science and Engineering,
Hiroshima University, Hiroshima 739-8526, Japan}


\begin{abstract}

The pulsar magnetosphere is a potential source of continuous gravitational waves due to the rapid charge-discharge process in short timescale, varying the electric-field energy density. We estimate the strain of the continuous gravitational waves, considering relativistic effects and different gap regions.
We find that the strain from the polar cap is too small, in contrast to previous results. 
On the other hand, the strain from the outer gap is as large as $\sim2\times10^{-24}$, enough for future gravitational-wave detection such as the Einstein Telescope. Our result presents a new approach for the future detection of gravitational waves to probe the physics in the magnetosphere.




\end{abstract}

\maketitle

\tableofcontents
%
%
%
%
%
%
%
%
%
\section{Introduction}

Gravitational waves are a unique tool for exploring compact objects in the universe. Gravitational wave interferometers have already detected merger events of binary black holes and binary neutron stars~\cite{KAGRA:2021vkt}. 
The masses of the detected black holes range from $\mathcal{O}(10)M_{\odot}$ to $\mathcal{O}(100)M_{\odot}$, and their distribution provides invaluable information about their 
origin~\cite{1964ApJS....9..201F,Barkat:1967zz,Woosley:2021xba,Sasaki:2016jop}. Gravitational wave observations also allow us to probe various physics, such as the testing of general 
relativity~\cite{LIGOScientific:2021sio}, and strong interaction under extreme conditions, which leads to constraints on neutron-star equation of states~\cite{LIGOScientific:2018cki}.
Such gravitational-wave astronomy has developed thanks to the discovery of gravitational waves emitted from the mergers of compact objects.


Not only such transient gravitational waves but also the continuous gravitational waves have been paid attention for, particularly from the most promising candidate, the neutron star. For instance, isolated neutron stars emit continuous gravitational waves at frequencies related to their rotation, typically at twice the spin frequency for quadrupolar deformations. This emission can arise from various types of deformations (e.g., elastic, magnetic, or magneto-thermal), and gravitational wave observations can therefore provide constraints on the internal properties of neutron stars (see \cite{2023LRR....26....3R} for a review).
Although current gravitational wave observations have not yet detected the continuous signals~\cite{KAGRA:2022dwb,LVK25}, future detectors such as the Einstein Telescope may be able to detect them~\cite{2025arXiv250312263A}. 

While internal deformations of neutron-star crust, such as crustal mountains or magnetic stresses in the stellar interior, have long been considered promising sources of continuous gravitational waves (\cite{2024CQGra..41d3001G} for a recent review), several studies have increasingly emphasized that the external magnetosphere itself can also contribute as the additional energy density. Early theoretical work established that anisotropic currents and plasma distributions in the magnetosphere can lead to time-varying quadrupole moments and hence to gravitational radiation beyond the stellar surface \cite{1996A&A...312..675B,2001A&A...367..525P,2003A&A...403..419L}. In recent studies, magnetosphere-driven scenarios have been revisited with modern plasma and general relativistic treatments. Analyses have demonstrated how current sheets, twisted magnetospheric fields, and global oscillations can amplify gravitational wave emission, sometimes at harmonics distinct from the stellar spin frequency~\cite{2020MNRAS.498..110N,2025arXiv250714634S}. 
As one such example of gravitational-wave emission, Ref.~\cite{Kouvaris25} has studied gravitational wave production in the so-called polar-cap model~\cite{1975ApJ...196...51R}, which arises from the periodic generation and disappearance of electric fields associated with gap formation and disruption at the polar caps.
They highlight that the magnetosphere, long recognized as central to neutron-star spin-down and high-energy emission, may also serve as an external "antenna" for gravitational radiation. Incorporating such contributions are essential for a comprehensive picture of neutron-star gravitational-wave phenomenology and for interpreting potential detections with present and future observatories.


In this work, we estimate the continuous gravitational waves produced by the rapid discharge cycles in the pulsar magnetosphere, considering a different accelerator, the outer gap, in addition 
to revisiting the polar-cap model. We finally verify the future gravitational-wave detectability in this scenario, i.e., whether the gravitational waves produced by the local time variation of the pulsar magnetosphere can be used as a probe of the particle acceleration mechanism. 


\section{Gravitational wave generation from a gap}

In the pulsar magnetosphere, the accelerating electric field undergoes periodic oscillations. Within the so-called "gap" regions, plasma depletion occurs when the current flowing through the magnetosphere cannot be sustained. This leads to the development of an electric field that accelerates charged particles. The accelerated particles emit high-energy gamma rays, which are converted to electron-positron pairs. These newly created pairs try to screen the electric field, thereby halting further particle acceleration and pair creation. However, in order to maintain the magnetospheric current, continuous particle injection is required. Once particle injection ceases, plasma depletion occurs again, leading to the redevelopment of the accelerating electric field. Through such a process, the oscillating behavior of the electric field is expected in the pulsar magnetosphere. Indeed, such behavior has been observed in plasma particle simulations, particularly in regions such as the polar cap or the outer region of the magnetosphere \cite{2007MNRAS.376.1460W, 2010MNRAS.408.2092T, 2023ApJ...958L...9B}.

We evaluate the amplitude of gravitational waves from a gap by correctly taking into account potential relativistic effects, which were neglected in previous works.
We consider a gap region where time-dependent electric fields exist along $z$-axis.
The spatial components of the energy-momentum tensor arising from the time-dependent electric field are given by
\begin{equation}
  T_{xx}(\bm{x},t) = T_{yy}(\bm{x},t) = -T_{zz}(\bm{x},t) = \frac{1}{2}E_{z}^{2}(\bm{x},t).
\end{equation}
Note that any time-independent electromagnetic fields are neglected, since they do not 
contribute to gravitational wave production.
The gravitational waves at a large distance from the source is given by~\cite{Maggiore:2007ulw}
\begin{equation}
  h_{ij}(t,\bm{x}) = \frac{1}{2\pi M_{pl}}\frac{1}{d} \Lambda_{ij,kl}(\hat{\bm{x}})
           \int^{\infty}_{-\infty} 
           \frac{d\omega}{2\pi} \tilde{T}_{kl}(\omega, \omega\hat{\bm{x}}) e^{-i\omega(t-r)} ,
           \label{GW}
\end{equation}
where $M_{pl}$ denotes the reduced Planck mass, and $\Lambda_{ij,kl}(\hat{\bm{x}})$ is the projection operator for the transverse-traceless gauge.
The Fourier transform of the energy-momentum tensor is defined as
\begin{equation}
  \tilde{T}_{kl}(\omega, \bm{k}) = \int d^{4}x\, T_{kl}(t,\bm{x}) e^{-i\omega t + i \bm{k}\cdot\bm{x}} .
  \label{Ttilde}
\end{equation}
We place the origin of the coordinate system at the source, so that $r$ represents the distance from the source to the observation point, i.e., $\bm{x}=r\hat{\bm{x}}$.
In general, evaluating Eq.\,(\ref{GW}) analytically is nontrivial.
Therefore, it is useful to define the gravitational wave flux, which represents the energy carried by gravitational waves per unit time and per unit area:
%
%
%
\begin{eqnarray}
  F &=& \frac{M_{pl}^{2}}{2}<\dot{h}_{ij}\dot{h_{ij}}> \label{Fh} \\
    &=& \frac{1}{16\pi^{3}M_{pl}^{2}d^{2}}\Lambda_{ij,kl}(\hat{x}) 
      \int^{\infty}_{0} d\omega \omega^{2} \tilde{T}_{ij}(\omega, \hat{\bm{x}}) 
      \tilde{T}^{*}_{kl}(\omega, \hat{\bm{x}}) \label{ene} .
\end{eqnarray}
The gravitational waves generated from the gap exhibit angular dependence with respect to the zenith angle $\theta$.
In fact, no gravitational waves are emitted along the $z$-axis, while the emission would be maximal at 
$\theta = \pi/2$.

\section{Gap configurations}

In this section, we investigate gravitational wave generation in two gap configurations using the result obtained in the previous section.
The first is the polar gap, and the second is the gap in the outer region of the magnetosphere.

\subsection{Polar gaps}

We model the polar gap as a rectangular box with side length $r_p$ and height $0<z<h$ for simplicity. 
The width $r_p$ is given by
\begin{equation}
  r_{p} \simeq R^{3/2}\Omega^{1/2} ,
\end{equation}
where $R$ is the radius of the pulsar, and $\Omega$ is the angular frequency of 
rotation of the neutron star.
The maximum height of the gap $h$ corresponds to the pair formation front where the curvature photon emitted from an accelerating electron or positron converts to pairs through the magnetic pair creation, and can be estimated as~\cite{1975ApJ...196...51R}
\begin{equation}
  h \simeq 5\times 10^{3}{\rm cm} \,  \left(\frac{\rho}{12{\rm km}}\right)^{2/7} 
                          \left(\frac{P}{1{\rm s}}\right)^{3/7}
                          \left(\frac{B_{s}}{10^{12}{\rm G}}\right)^{-4/7} , 
\end{equation}
where $\rho$ is the curvature radius of the magnetic field, and $P(=2\pi/\Omega)$ is the rotational period, and $B_s$ is the strength of the magnetic field at the surface. 
We assume that the electric field $E$ in the polar cap periodically varies with a characteristic time scale $T$.  
We adopt the following model to represent the temporal variation of the electric field,
\begin{equation}
  E(\bm{x},t) =
\begin{cases}
    - E_{{\rm max}} \dfrac{h - z  }{h} \, \theta(z(t)-z), 
       \quad z(t) = \beta \left( t + \frac{T}{2} \right) 
      \quad & \text{for } -\frac{T}{2} \leq t \leq -\frac{T}{2} + \frac{h}{\beta}, \\[10pt]
    - E_{{\rm max}} \dfrac{h - z  }{h}, 
      & \text{for } -\frac{T}{2} + \frac{h}{\beta} \leq t \leq \frac{T}{2} - \frac{h}{\beta}, \\[10pt]
    - E_{{\rm max}} \dfrac{h - z  }{h} \, \theta(z(t)-z), 
       \quad z(t) = \beta \left(-t + \frac{T}{2} \right) 
      \quad & \text{for }\,\quad \frac{T}{2} - \frac{h}{\beta} \leq t \leq \frac{T}{2} ,
\end{cases} \label{E}
\end{equation}
where $E_{\max}$ represents the maximum amplitude of the electric field, $\beta$ denotes the velocity of the gap front, and $\theta(z)$ is the step function.
Note that the electric field model considered here, which depends linearly on $z$, differs from that in~\cite{Kouvaris:2024ait}, where the electric field was assumed to be time-dependent but spatially 
homogeneous. However, it only affects the result by an overall numerical factor.
The maximum amplitude of the electric field is given by
\begin{equation}
    E_{\max}=2\Omega B_s h.
\end{equation}
The electric field is directed along the $z$-axis, which is parallel to the magnetic field, such that $z = 0$ corresponds to the inner boundary of the gap. 
In this model, there is no gap at $t = -T/2$.
During the interval $-T/2 \leq t \leq -T/2 + h$, the outer boundary of the gap moves at a velocity $\beta$ which
can be relativistic.
After this phase, the gap height remains constant until $t = T/2 - h$, and then begins to decrease over the interval $T/2 - h \leq t \leq T/2$, also at velocity $\beta$.

Using Eqs. (\ref{ene}) and (\ref{E}), one can evaluate the energy flux of gravitational waves from the polar gap as
\begin{equation}
  F \simeq 
    \frac{h V E_{{\rm max}}^4}{288\pi^2 M_{pl}^{2} r^{2}}, \label{flux} 
\end{equation}
where $V$ 
denotes the volume of the gap. 
The direction of the emitted gravitational waves is taken to be perpendicular/parallel to the side faces of the rectangular box for simplicity.
We also averaged over the $\theta$-dependence, approximating it as $\sin\theta$, which agrees well with numerical calculations.
Comparing this expression with Eq.\,(\ref{Fh}), 
we estimate that the amplitude of the gravitational waves at the characteristic frequency 
$\omega/2\pi = 2/ T$, which corresponds to twice the frequency of the gap cycle, is given by
\begin{eqnarray}
 h_{0} &\sim&    \frac{ h^{1/2} V^{1/2} T E_{{\rm max}}^{2} }
                     {48 \pi^{2}  M_{pl}^{2}  r} .  \label{h0}  
\end{eqnarray}
Typically, $T \sim \mathcal{O}(10)\times h$.
Therefore, from Eq.\,(\ref{h0}), we can estimate the amplitude of the gravitational waves emitted from the polar cap:
\begin{equation}
 h_{0}  \sim  9.5 \times 10^{-47} \left(\frac{B_{s}}{10^{16}{\rm G}}\right)^{\frac{2}{7}} 
                              \left(\frac{10^{-3}{\rm s}}{P}\right)^{\frac{17}{14}} 
                              \left(\frac{R}{12{\rm km}}\right)^{\frac{3}{2}}
                              \left(\frac{\rho}{12{\rm km}}\right)^{\frac{6}{7}}
                              \left(\frac{T}{1.1\times 10^{-7}{\rm s}}\right)
                              \left(\frac{1{\rm kpc}}{r}\right) . 
                              \label{h02}
\end{equation}
We mention that the estimated gravitational wave amplitude is much smaller than that reported in the previous work~\cite{Kouvaris25}.
Their calculation is based on the non-relativistic approximation, which is valid when the source velocity is much smaller than the speed of light, whereas our analysis does not rely on this assumption, as the velocity associated with the moving gap height can be relativistic.
If one incorrectly adopts the non-relativistic approximation, the result is overestimated by a factor of $\sim \omega r_p$, because the term $\bm{k}\cdot\bm{x}$ in the exponent of Eq.(\ref{Ttilde}) is neglected in the approximation.
Moreover, while Eq.~(\ref{h02}) does not explicitly depend on the stellar radius $R$, 
the expression in Ref.,\cite{Kouvaris25} does, which could result in further overestimation.
From Eq.\,(\ref{h02}), one can see that the amplitude of the generated gravitational waves is not particularly sensitive to neutron star parameters such as the magnetic field strength or the rotation period.
The characteristic frequency, given by $\omega/2\pi = 2/T$, lies in the high-frequency regime, 
typically around 10MHz.
Although the amplitude of these gravitational waves is far below the sensitivity of current high-frequency gravitational wave observations,
this result encourages us to develop new detection methods for high-frequency gravitational 
waves~\cite{Aggarwal:2025noe, Ito:2025mgm, Matsuo:2025blj, Ito:2019wcb, Ito:2020wxi, Berlin:2021txa, Ito:2022rxn, Ito:2023bnu, Berlin:2023grv, Bringmann:2023gba, Kanno:2023whr, Ejlli:2019bqj, Domcke:2022rgu,Ito:2023nkq,Ito:2023fcr}.

\subsection{Gaps in outer region of magnetosphere}
There may also exist gaps in the outer region of the magnetosphere, whose distance from the center of the neutron star is roughly given by the light cylinder radius, $\sim R_{{\rm lc}} (= \Omega^{-1})$. 
Such a gap may experience cycles of the gap formation and disruption 
as same as the case of the polar gap discussed in the previous section.
Therefore, we expect gravitational wave emission from these gaps in outer magnetosphere as well.

We model such a gap as a rectangular box, 
where the $z$-direction is taken to be aligned with the dipole magnetic field, the
$y$-direction corresponds to the azimuthal axes, 
and the $x$-direction is orthogonal to them.
The side lengths of the gap are denoted by $h_z$, $h_y$, and $h_x$, respectively.
In the gap, the amplitude of dipole magnetic fields is $B_{{\rm out}}\sim B_{s}(R/ R_{{\rm lc}})^{3}$.
Let us consider a simple electric fields model in the gap: 
\begin{equation}
  \tilde{E}(\bm{x},t) =
\begin{cases}
     \tilde{E} \, \theta(x(t)-x), 
       \quad x(t) = \beta \left( t + \frac{T}{2} \right) ,
      \quad & \text{for } -\frac{T}{2} \leq t \leq -\frac{T}{2} + \frac{h_{x}}{\beta}, \\[10pt]
    \tilde{E} , 
      & \text{for } -\frac{T}{2} + \frac{h_{x}}{\beta} \leq t \leq \frac{T}{2} - \frac{h_{x}}{\beta}, \\[10pt]
    \tilde{E}  \,  \theta(x(t)-x), 
       \quad x(t) = \beta \left(-t + \frac{T}{2} \right) ,
      \quad & \text{for }\,\quad \frac{T}{2} - \frac{h_{x}}{\beta} \leq t \leq \frac{T}{2} .
\end{cases} \label{Eout}
\end{equation}
where $z=0$ and $x=0$ correspond to the inner boundary of the gap in the $z$ and $x$ directions, respectively.
$\tilde{E} = \Omega B_s h_x \alpha (R/ R_{{\rm lc}})^{3}$ represents the 
typical amplitude of the electric fields,
where $\alpha$ represents a geometrical factor related to such as the boundary condition of the gap, 
for example,
$\alpha \sim \left( \frac{h_x}{R_{{\rm lc}}}\right)$ for the outer gap model~(see Eq. (3.10) in \cite{1986ApJ...300..500C}).
For simplicity, the electric field inside the gap is assumed to be homogeneous.
The gap height $h_x$ is likely determined by the mean free path of photons in the gap, 
and can be estimated as $h_x \sim \mathcal{O}(0.1)\, R_{\rm lc}$.

Using Eqs.,(\ref{ene}) and (\ref{Eout}), one can evaluate the energy flux of gravitational waves
from the gap as
\begin{equation}
  F \simeq 
    \frac{\beta \gamma^2 V^2 \tilde{E}^4}{15 \pi^3 M_{pl}^{2} h_x T r^{2}},
    \label{fluxout} 
\end{equation}
where $V=h_xh_yh_z$ represents the volume of the gap.
In the above expression, we have averaged over the $\theta$-dependence by approximating it with $\sin^4\theta$, which shows good agreement with numerical calculations.
Comparing this expression with Eq.\,(\ref{Fh}), 
we estimate that the amplitude of the gravitational waves at the characteristic frequency 
$\omega/2\pi = 2/T$, which corresponds to twice the frequency of the gap cycle, is given by
\begin{eqnarray}
 h_{0} &\sim&   \frac{\gamma  V  \beta^{1/2} T^{1/2} \tilde{E}^{2} }
                     {2 \sqrt{30} \pi^{5/2} M_{pl}^{2} h_x^{1/2} r} .  \label{h0out}  
\end{eqnarray}
One can see that there is a relativistic enhancement $\gamma$, in contrast to the polar gap case 
where the directions of the gap oscillation and the electric field component are parallel.
Assuming that the period of the gap formation and disruption is $T \sim \kappa \beta^{-1}h_{x}$, and 
$h_y=h_z\sim R_{{\rm lc}}$,
one can estimate the gravitational wave amplitude from the gap in the outer magnetosphere:
\begin{eqnarray}
 h_{0} &\sim& 
      2.4 \times 10^{-24}     \left(\frac{\gamma}{10}\right)
                              \left(\frac{B_{s}}{10^{16}{\rm G}}\right)^{2} 
                              \left(\frac{10^{-3}{\rm s}}{P}\right)^{3} 
                              \left(\frac{R}{12{\rm km}}\right)^{6}
                              \left(\frac{\kappa}{10}\right)^{1/2}
                              \left(\frac{\alpha}{1}\right)^{2}
                              \left(\frac{1{\rm kpc}}{r}\right) ,  \label{h03}
\end{eqnarray}
at the frequency of $\omega/2\pi = 2/T \simeq 13 {\rm kHz}$.
\begin{figure}[ht]
\centering
\includegraphics[width=9cm]{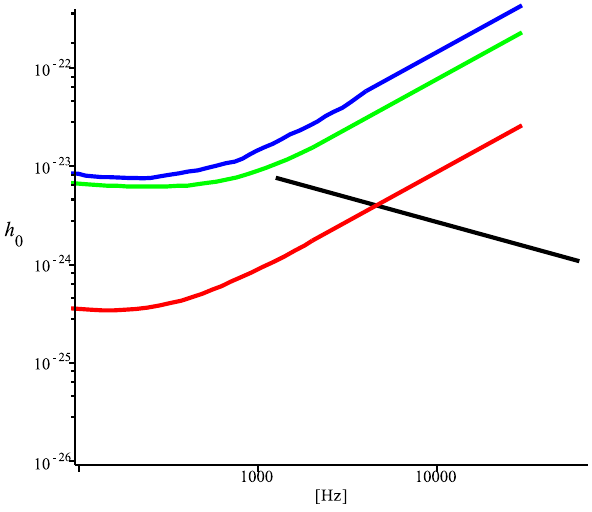}
\caption{The gravitational wave amplitude $h_0$ from the outer gap region is shown.
The black line represents $h_0$ calculated from Eq.~(\ref{h03}) with $\kappa$ varied from $2$ to $100$.
The blue, green, and red curves show the sensitivities of the combined LIGO O2 and O3 runs~\cite{LIGOScientific:2021hvc},
the combined LIGO O4 and future O5 runs~\cite{Riles:2022wwz},
and the Einstein Telescope~\cite{Riles:2022wwz}, respectively. 
} 
\label{figh0}
\end{figure}
In Fig. \ref{figh0}, we compare the predicted $h_0$ of Eq.~(\ref{h03}), with $\kappa$ varied from $2$ to $100$,
to the estimated sensitivities of
the combined LIGO O2 and O3 runs~\cite{LIGOScientific:2021hvc},
the combined LIGO O4 and future O5 runs~\cite{Riles:2022wwz},
and the Einstein Telescope~\cite{Riles:2022wwz}. 
We also take into account the spin-down timescale, which is given as:
\begin{eqnarray}
    t_{\rm sd} = 22~{\rm s} \left(\frac{I}{1.61\times10^{45}~\mathrm{g~cm^2}}\right)\left(\frac{B_s}{10^{16}~\mathrm{G}}\right)^{-2}\left(\frac{R}{12~\mathrm{km}}\right)^{-6}\left(\frac{P}{10^{-3}s}
    \right)^2~,
\end{eqnarray}
where $I$ is the moment of inertia \cite{2019RPPh...82j6901E}, and the observation time used to evaluate the sensitivities 
is assumed to be $t_{\rm sd}$. 
Despite such a short exposure time, the strain could reach the sensitivity of 
the Einstein Telescope, implying future detectability.

\section{Discussion}

In this paper, we calculated the continuous gravitational waves produced by the rapid discharge cycles in the pulsar magnetosphere, considering relativistic effects and different gap regions. We found that the strain from the polar cap is too weak to be detected.
On the other hand, the strain from the outer gap can be around $2.4\times10^{-24}$ assuming a source distance of 1 kpc, which is sufficient for future gravitational-wave detection as shown in Fig. \ref{figh0}.
The gravitational wave signal sourced by time-dependent electromagnetic energy density
in the outer magnetosphere naturally appears at ${\rm kHz}$ frequencies. 

Note that even if there is no detection of continuous gravitational waves in future observations, 
through the constraint on the upper limit of electric-field energy, one can probe the magnetosphere physics, possibly including the non-linear plasma effects beyond standard polar/outer cap models, such as the intermittent screening of the accelerating electric field and magnetic reconnection due to current sheet formation (for reviews, see e.g., \cite{2016JPlPh..82e6302P,2017SSRv..207..111C,2022ARA&A..60..495P}). Considering such a dynamical magnetosphere in the hunting of continuous gravitational waves is left for our future work.

\begin{acknowledgements}
This work was in part supported by JSPS KAKENHI Grant Number 22K14034 (A. I.), 22K03681, 23K22538 (S. K.), 23K19056 and 25K17403 (A. D.).
\end{acknowledgements}

\bibliography{ref}

\end{document}